\documentstyle[12pt,epsfig]{article}

\begin{document}

\begin{center}

{\large \bf  Exactly Solvable Model for the QCD Tricritcal Endpoint}
%

\vspace{1.0cm}

{\bf Kyrill A. Bugaev
}\\

\vspace{1.cm}

\vspace{0.5cm}

Bogolyubov Institute for Theoretical Physics,\\
03680 -- Kiev, Ukraine\\

\date{\today}

\end{center}

%

%
\begin{abstract}
An inclusion of  temperature and  chemical potential  dependent  surface tension
 into  the gas of quark-gluon bags model  resolves a long standing problem  of  a unified description of the first and second order phase transition with the cross-over. 
The suggested model has an exact analytical solution and allows one to rigorously study 
the vicinity of the critical endpoint  of the deconfinement  phase transition. 
It is  found that at the curve  of  a zero surface tension coefficient 
 there must exist  the surface induced  phase tranition of the 2$^{nd}$ or higher order.
The present  model predicts that the  critical endpoint (CEP)  of quantum chromodynamics  is the tricritical endpoint.
\end{abstract}
%
%

\newpage

\section{Introduction}

The role of surface tension for the quark gluon plasma (QGP) was discussed long ago  \cite{Jaffe:1, Jaffe:2}, however, up to now the situation is  somewhat  unclear 
\cite{Bugaev:07,Moretto:0511}.  In nuclear and cluster physics 
the importance of the surface tension for the properties of CEP  
is known from a number of  exactly solvable cluster  models with 
the 1$^{st}$ order phase transition (PT)  which describe the critical point properties  very well.
These models are built on the assumptions that 
the difference of  the bulk  part (or the volume dependent part) of  free energy  
of two phases disappears at  phase equilibrium and that, in addition, 
the difference of the surface part (or the surface tension) of  free energy  vanishes 
at the critical point. 
The most famous of them is  the Fisher droplet model (FDM) \cite{Fisher:67,Elliott:06}
which has been successfully used to analyze the condensation  of a gaseous phase 
(droplets of all sizes)   into a liquid. 

Another such a model is  a simplified version of  the  statistical multifragmentation  model (SMM)  \cite{simpleSMM:1} which was 
solved analytically both for infinite \cite{Bugaev:00,Reuter:01} and 
for finite \cite{Bugaev:04a, Bugaev:05c} volumes of the system. 
The analysis of critical indices of the SMM  \cite{Reuter:01} shows that 
the value of   Fisher exponent  $\tau_{SMM} = 1.825 \pm 0.025$ of  this 
model  is consistent with ISiS Collaboration data  \cite{ISIS}
and EOS Collaboration data  \cite{EOS:00}. 
Such an   experimentally obtained range of the $\tau$ index is of a principal importance because  
it gives a very strong evidence that the SMM, and, thus, 
the nuclear matter,  has a tricritical endpoint rather than a critical  endpoint 
\cite{Bugaev:00,Reuter:01}. 

This  success of  the SMM initiated  
the studies of the surface partitions of large clusters  
within the Hills and Dales Model
 \cite{Bugaev:04b,Bugaev:05a} and led to a discovery of the origin  of
the temperature independent surface entropy similar to the FDM.  
As a consequence, the surface tension coefficient of large 
clusters consisting of the discrete constituents should linearly depend 
on the temperature of the system \cite{Bugaev:04b} and  must vanish at the critical endpoint.
However, the present formulation of the Hills and Dales Model
\cite{Bugaev:04b,Bugaev:05a}, which successfully 
estimates the upper and lower bounds of the surface deformations of the discrete 
physical clusters, does not look  suitable for  quark-gluon bags.
Therefore, in this work I  insert the surface tension into the gas of bags model 
(GBM)  \cite{Goren:81},  assume a certain  dependence of the surface 
tension coefficient on temperature and baryonic chemical potential.
Then I analyze the quark gluon bags with surface tension (QGBST) model  and  
concentrate  on the impact  of  surface tension  on  the properties of 
the deconfinement  phase diagram and  the QCD critical endpoint.

Here I show  that  at  low values of the  baryonic chemical potential
the 1$^{st}$ order deconfinement PT degenerates into a cross-over, if 
the surface tension coefficient becomes negative for lower values of temperature
than  the  transition  temperature. 
Also I prove   the existence of an additional PT 
of the 2$^{nd}$ or higher order  along the curve where the surface tension coefficient  vanishes. 
Thus, I am arguing that the QGBST model predicts 
the existence of the tricritical rather than critical endpoint. 

\section{The Role of Surface Tension at Zero Baryonic Densities}

The isobaric partition of the QGBST model obtained from the grand canonical one $Z(V,T)$ is as follows
\begin{eqnarray}\label{Zs}
& \hat{Z}(s,T) \equiv \int\limits_0^{\infty}dV\exp(-sV)~Z(V,T) =\frac{1}{ [ s - F(s, T) ] }   \,.
\end{eqnarray}
Here the function $F(s, T)$  consists of two parts, the discrete mass-volume spectrum $ F_H(s,T)$, and 
the continuous part of the spectrum $ F_Q(s,T)$
\begin{eqnarray}
F(s,T)&\equiv  \sum_{j=1}^n~  \phi(T,m_j)  e^{-v_js}
+ ~ u(T)
 ~  \int\limits_{V_0}^{\infty}dv~ \frac{ \exp\left[ \left( s_Q(T)-s \right) v - \sigma(T)\, 
v^{\kappa} \right] }{v^{\tau}} 
~,
 \label{FQs}
\end{eqnarray}
where  the  function $ \phi(T,m_k) \equiv  \frac{g_k}{2\pi^2} \int\limits_0^{\infty}\hspace*{-0.1cm}p^2dp~
e^{\textstyle - \frac{(p^2~+~m_k^2)^{1/2}}{T} }
=  g_k \frac{m_k^2T}{2\pi^2}~{ K}_2\left( \frac{m_k}{T} \right)
$ 
is the particle  density
of  bags of mass $m_k$ and eigen volume $v_k$  and degeneracy $g_k$.

At the moment the particular choice of function $F_Q(s,T)$ in (\ref{FQs}) is not important. 
The key point of my treatment  is that it should have the form of Eq.  (\ref{FQs}) which 
has a singularity  at  $s=s_Q^*$ 
because for $s<s_Q$ the integral over the bag   volume $v$  diverges at its upper limit. 
As will be shown below the isobaric partition (\ref{Zs}) has two kind of singularities: the simple pole $s = s_H^*$ and the essential singularity
$s=s_Q$  The rightmost singularity defines the phase in which matter exists,
whereas a PT occurs when two singularities coincide \cite{Goren:81,Bugaev:00,  Bugaev:07}. 
All singularities are defined by the equation
\begin{eqnarray}\label{s*vdw}
  s^* ~ &= & ~   {F (s^*, T)}\,.
 \end{eqnarray}

The $v$-linear term in 
the exponential of  the continuous spectrum (\ref{FQs}) is nothing else, but a difference of the bulk 
free energy of a bag of volume $v$, i.e. $ -T s v$,  which is  under external pressure 
$T s$,   and  the bulk  free energy of 
the same bag filled with QGP, i.e.  $ -T s_Q v$. 
The term  $- T s v$ appears due to the hard core repulsion \cite{Bugaev:07}, whereas the  QGP pressure, 
$T s_Q$,   appears in  (\ref{FQs}) as a generalization of the  Hagedorn  mass spectrum \cite{Hagedorn:65}. 
At phase equilibrium this difference of the bulk free energies  vanishes and the properties of phase 
equilibrium  are defined  by the surface free energy. 

Note that the usage of the grand canonical  description for  the exponential mass or volume spectrum  
of Hagedorn type 
was  strongly criticized  recently \cite{Moretto:0511,Moretto:06,Moretto:06b,BugaevMoretto:05a,Elliott:05}  because  of the thermostatic properties of this spectrum. However, the hard core repulsion  compensates  the growing part of the mass-volume spectrum and, hence, the criticism of Refs.  \cite{Moretto:0511,Moretto:06,Moretto:06b,BugaevMoretto:05a,Elliott:05} is irrelevant to the present model.

The new element in  (\ref{FQs})  is the presence of surface free energy  
$\sigma_0 v^{\kappa}$ (${\kappa}< 1$) of the bag. The power  $\kappa < 1$ which describes the bag's effective  surface is a constant which,  in principle, can differ from the typical FDM and SMM value  $\frac{2}{3}$, 
if  the highly non-sperical  bags are
possible \cite{Bugaev:07,Elliott:06,Bugaev:04b,Bugaev:05a}. 
The ratio of the  temperature dependent surface tension coefficient  to $T$
(the reduced surface tension coefficient hereafter) 
which has the form $\sigma(T) = 
\frac{\sigma_o}{T} \cdot
\left[ \frac{ T_{cep}   - T }{T_{cep}} \right]^{2k + 1} $  ($k =0, 1, 2,...$).  
Here $\sigma_o > 0$ can be a smooth function of the temperature, but for simplicity I  fix it to be a constant.  

In choosing such a simple surface energy parameterization I 
follow the original Fisher idea \cite{Fisher:67}  which allows one to account for 
the surface energy by considering 
some mean bag of volume $v$ and surface $v^{\kappa}$. The consideration of 
the general mass-volume-surface bag spectrum is reserved  for the future investigation. 
In contrast to the FDM and SMM,  the power  $\kappa < 1$ which describes the bag's effective  surface is a constant which,  as mentioned above, can differ from the typical FDM and SMM value $\frac{2}{3}$.
This is so because  near  the deconfinement PT region  the QGP  has  low density and, hence, 
like in the low density  nuclear matter \cite{Ravenhall},  
the non-sperical bags (spaghetti-like or lasagna-like \cite{Ravenhall})  can be  favorable 
(see a \cite{Bugaev:07} and references therein). 
A similar idea of  ``polymerization" of gluonic quasiparticles was introduced recently 
\cite{Shuryak:05a}. 

The second  essential  difference with the FDM and SMM surface tension 
parameterizations is that   the  vanishing of $\sigma(T)$ above the CEP temperature  is not required. 
As will be shown later,  this is the most important assumption which, in contrast to the GBM,  allows one to naturally describe the cross-over  from hadron gas to QGP.  
Note that  negative value of the reduced surface tension coefficient $\sigma(T)$ above the CEP 
does not mean anything wrong. As  discussed above, 
the  surface  tension coefficient consists of energy and entropy parts which  have   opposite signs \cite{Fisher:67,Bugaev:04b,Bugaev:05a}. 
Therefore, $\sigma(T) < 0 $ does not mean that the surface energy changes the sign, but it
rather  means that the surface entropy, i.e. the logarithm of the degeneracy of bags of a fixed volume, simply  exceeds their  surface energy.  In other words, 
the number of  non-spherical bags of a fixed volume becomes so large that the Boltzmann exponent, which accounts for the energy "costs" of these bags,  cannot
suppress them anymore.  

Finally, the third essential difference with the FDM and SMM is that it is  assumed 
that the surface tension in the QGBST model vanishes  at some line in $\mu_B-T$ plane, 
i.e. $T_{cep} = T_{cep} (\mu_B)$. However,  in the subsequent  sections I will 
consider $T_{cep} = Const $ for simplicity, and in Sect. 4   I  will  discuss the necessary modifications of the model  with $T_{cep} = T_{cep} (\mu_B)$.

In principle, besides  the bulk and surface parts of  free energy,  the continuous   volume  spectrum $F_Q(s, T)$ could include the curvature part as well, which may be important for small hadronic bubbles \cite{Svet:1, Svet:2, Madsen:00} or for cosmological PT \cite{Ignat:1}.  It is necessary to stress, however,
that the critical properties of the present model are defined by the infinite bag, therefore
the inclusion into the function $F_Q(s, T)$ of a curvature term of bag's free energy  of any  sign could  affect 
the thermodynamic quantities of this  model at $s = s_Q(T)$ and $\sigma (T) = 0$,
which is possible  at (tri)critical  endpoint only (see below). 
If, the curvature term was really  important for the cluster models like the present one, then  it should have been seen also at  (tri)critical  points of  the FDM, SMM and many systems  described by the FDM, but this is not the case \cite{Elliott:06,Complement}. 
Indeed,  recently  the  Complement method   \cite{Complement} was applied to 
the analysis of the  largest, but still mesoscopic  drop of a radius $R_{dr}$  representing the liquid in equilibrium with its vapor.  The method allows one to find out  the concentrations 
of  the vapor clusters in finite system in a whole range of  temperatures and determine 
the free energy  difference of  two phases with high precision. The latter enables us 
not only to  extract  the critical temperature, surface tension coefficient and even the  value of Fisher index $\tau$ of the  infinite system,  but also such a delicate effects as 
the Gibbs-Thomson correction  \cite{Gibbs-Thomson}  to the free energy of a liquid 
drop.  Note that  the Gibbs-Thomson correction  behaves as   $R_{dr}^{-1}$, but the  Complement method   
\cite{Complement}  allows one to  find it, whereas the curvature part of free energy, which is proportional to 
$R_{dr}$,  is not seen  both for a drop and for smaller clusters. 
Such a result   is  directly related  to  the QGP bags because 
QCD  is expected to be in the same universality class \cite{Rob,misha} as 
the 3-dimensional Ising model whose clusters were analyzed  in \cite{Complement}. 
Therefore, admitting that for  finite QGP bags  the curvature effects may be essential,  
I leave them out  because the critical behavior of the present model is defined  
by the properties of  the  infinite bag.
On the other hand,  similarly  to  the FDM, SMM and 
FDM-like systems  discussed in Ref. \cite{Elliott:06},   
I assume that  the curvature part  of  free energy of  the  infinite QGP bag is 
not important and  leave  for  the future analysis the question why this is so.

According to the general theorem \cite{Goren:81} the analysis of PT  existence of the GCP  is now reduced to the analysis of the rightmost singularity of 
the isobaric partition (\ref{Zs}).  Depending on the sign of the reduced surface tension coefficient, there are three possibilities. \\

\noindent
({\bf I}) The first possibility corresponds to $\sigma(T) > 0$. Its treatment is  very similar 
to the GBM parameterization  of  continuous spectrum  with $\tau > 2$ \cite{Goren:81}. In this case at low 
temperatures the QGP  pressure $T s_Q(T)$ is negative and, therefore, the   rightmost singularity is a simple pole of the isobaric partition  
$s^* = s_H (T) = F(s_H(T), T) > s_Q(T)$, which is mainly defined by a discrete part of  the volume spectrum $F_H(s,T)$. 
The last inequality provides the convergence of the volume integral in (\ref{FQs})
(see Fig. 1). On the other hand at very high $T$ the QGP pressure dominates and, hence, the rightmost singularity is the essential  singularity  of the isobaric partition  $ s^* = s_Q(T)$.  
The phase transition occurs, when the singularities coincide:
\begin{equation}\label{PTI}
  s_H (T_c) \equiv  \frac{p_H (T_c)}{T_c} =  s_Q (T_c) \equiv  \frac{p_Q (T_c)}{T_c}\,,
 \end{equation}
which is nothing else, but the Gibbs criterion. 
The graphical solution of Eq. (\ref{s*vdw})  for  all  these possibilities is shown in Fig. 1.  
Like in the GBM \cite{Goren:81}, the  necessary condition for the PT existence is  the finiteness of $F_Q(s_Q(T), T) $ at $s = s_Q(T)$.
It can be shown that the sufficient conditions are   the following 
inequalities: $F_Q(s_Q(T), T) > s_Q(T) $ for low temperatures and 
  $F(s_Q(T), T) < s_Q(T) $  for  $T \rightarrow \infty$.  These 
conditions provide that at low $T$ the rightmost singularity  of the isobaric partition is a simple pole,
whereas for hight $T$ the essential singularity 
 $s_Q(T) $ becomes  its  rightmost one   (see Fig. 1 and a detailed 
analysis of the case $\mu_B \neq 0$).

%
%
\begin{figure}[ht]
%
\centerline{\hspace*{-0.0cm}\epsfig{figure=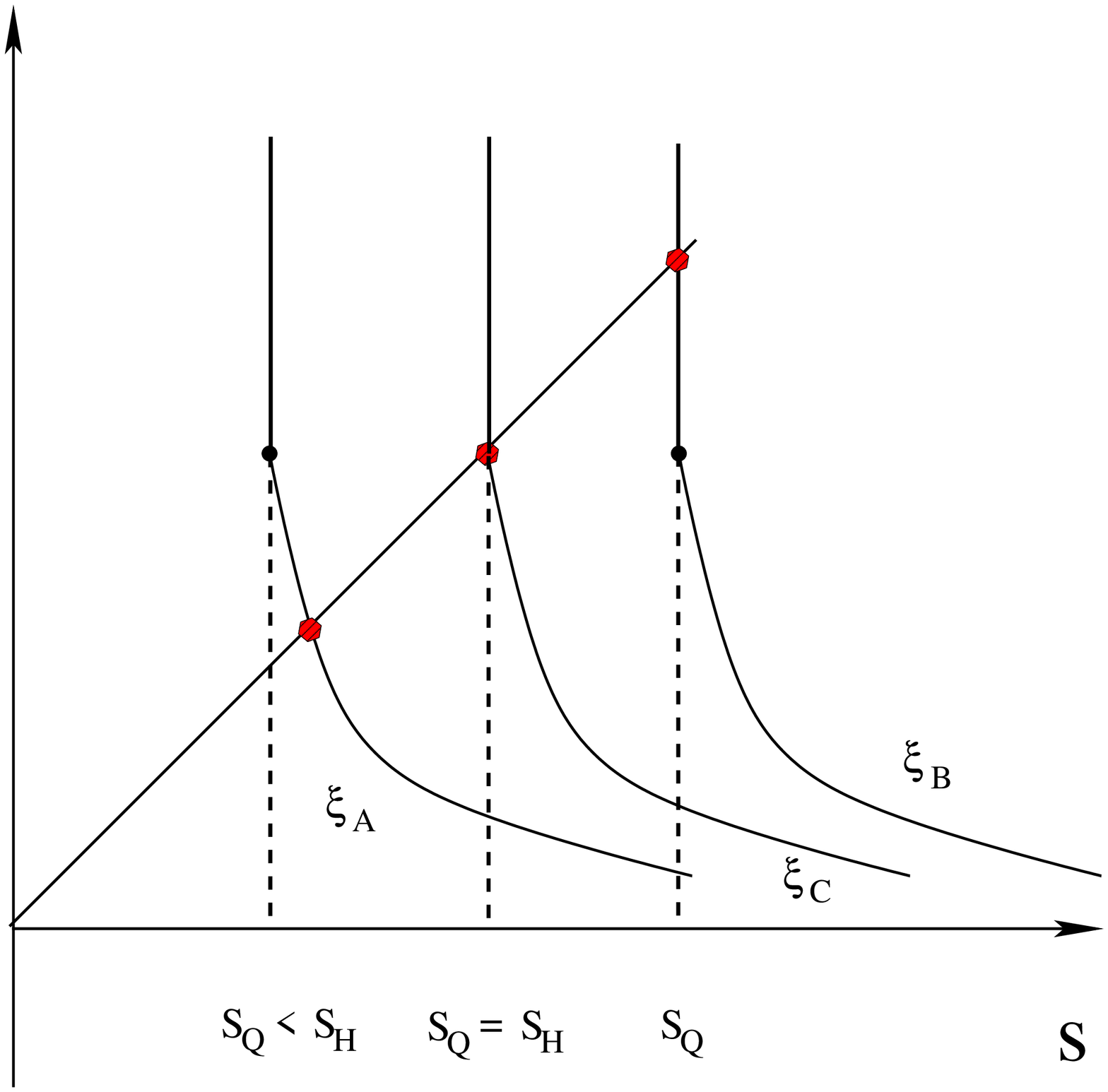,height=6.cm,width=6.0cm} 
\hspace*{1.0cm}  
\epsfig{figure=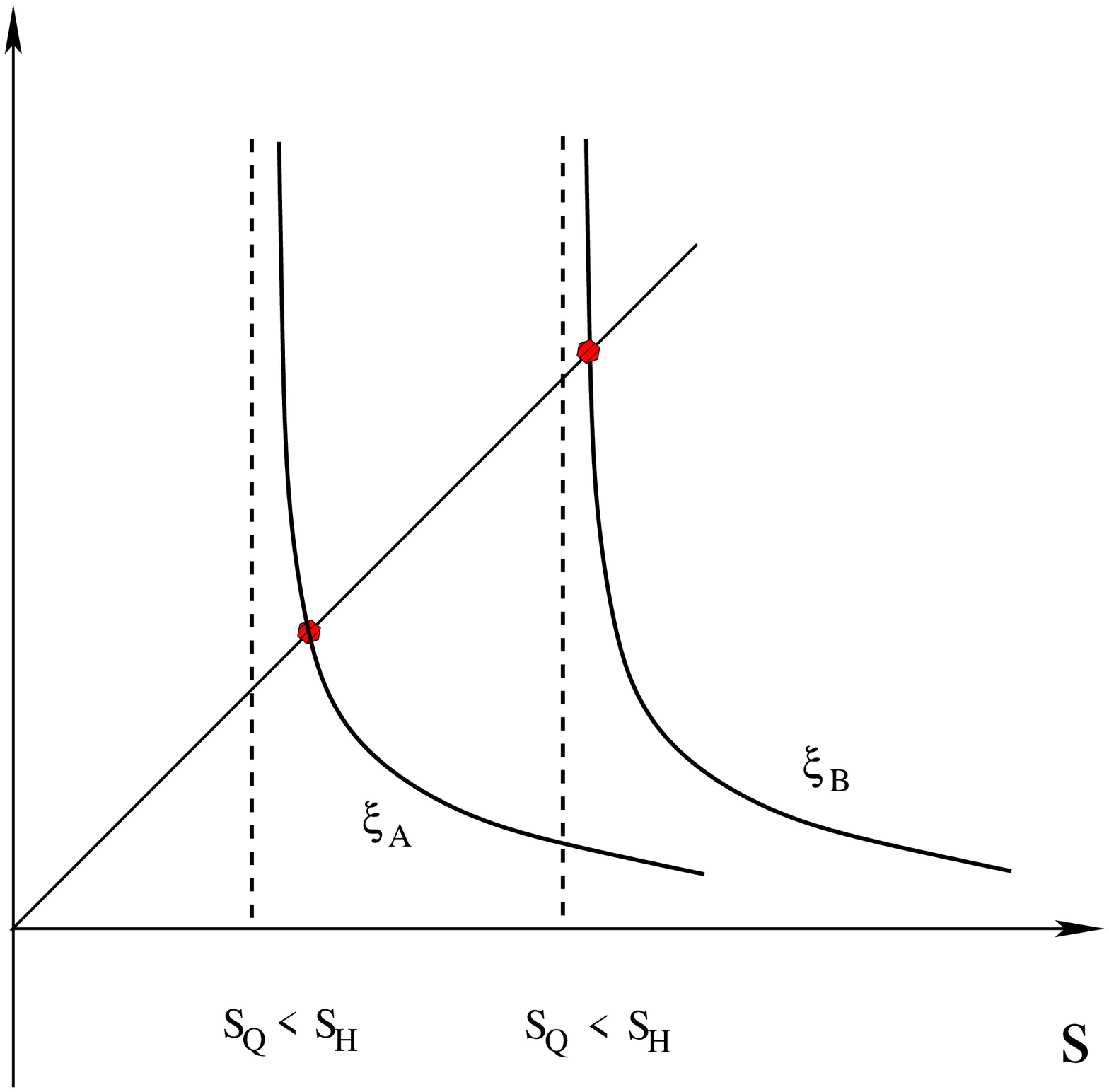,height=6.cm,width=6.0cm}}
\vspace*{-0.5cm}
\caption{
\label{fig1}
{\bf Left panel.}
Graphical solution of Eq. (\ref{s*vdw}) which corresponds to a PT.
The solution of Eq. (\ref{s*vdw}) is shown by a filled hexagon.
The function $F(s, \xi)$ is shown by a solid curve for a few 
values of  the parameter $\xi$.  The function  $F(s, \xi)$ diverges 
for $s < s_Q( \xi)$ (shown by dashed lines), but is finite at $s = s_Q( \xi)$ (shown by black circle).  
At low values of  the parameter $\xi = \xi_A$, which can be either $T$ or $\mu_B$, 
the simple pole $s_H$ is the rightmost singularity and it corresponds to hadronic phase. 
For  $\xi = \xi_B \gg \xi_A$ the  rightmost singularity is an essential singularity $s = s_Q( \xi_B)$, 
which describes QGP. 
At intermediate  value $\xi = \xi_C$ both singularities coincide $s_H( \xi_C) = s_Q( \xi_C)$ and 
this condition is a Gibbs criterion. 
\newline
{\bf Right panel.} Graphical solution of Eq. (\ref{s*vdw}) which corresponds to a cross-over.
The notations are the same as in the left panel. 
Now the function  $F(s, \xi)$ diverges 
at $s = s_Q( \xi)$ (shown by dashed lines). 
In this case the simple pole $s_H$ is the rightmost singularity for any value of $\xi $. 
}

\end{figure}

\vspace*{0.7cm}

The PT order can be found from the $T$-derivatives of  $s_H (T)$. 
Thus,  differentiating  (\ref{s*vdw}) one finds 
\vspace*{-0.2cm}
\begin{eqnarray}\label{sHprime1}
s_H^{\prime}~=~
\frac{G~+~u\,{\cal K}_{\tau-1}(\Delta, - \sigma) \cdot  s_Q^{\prime}}{1~+~u\,{\cal K}_{\tau-1}(\Delta, - \sigma)} \,,
\end{eqnarray}

\vspace*{-0.2cm}

\noindent
where the functions $G $ and ${\cal K}_{\tau -a} (\Delta, - \sigma) $ are defined as ($\Delta \equiv s_H - s_Q$)
\begin{eqnarray}\label{G1}
&\hspace*{-0.25cm}G \equiv F_H^{\prime}+ \frac{u^{\prime}}{u} F_Q 
+  {\textstyle\frac{ (T_{cep} - 2k T) \sigma(T)}{(T_{cep} - T)\, T } }\,u \,
{\cal K}_{\tau- \kappa}  (\Delta, - \sigma)\,, \\
\label{KQ}
&\hspace*{-0.25cm}{\cal K}_{\tau -a} (\Delta, - \sigma) \equiv  \hspace*{-0.0cm}
\int\limits_{V_o}^{\infty}\hspace*{-0.05cm}dv~\frac{\exp\left[-\Delta v - \sigma(T) 
v^{\kappa} 
\right] }{v^{\tau-a}} \,.
\end{eqnarray}

Now it is easy to see that the transition is of the 1$^{st}$ order,
i.e. $s_Q^{\prime}(T_c)>s_H^{\prime}(T_c)$, provided  $ \sigma(T) > 0$ for any $\tau$.
The 2$^{nd}$ or higher order phase transition takes place
provided $s_Q^{\prime}(T_c)=s_H^{\prime}(T_c)$ at $T=T_c$.
The latter condition is satisfied  when $ {\cal K}_{\tau-1}$ diverges to infinity
at $T\rightarrow (T_c-0)$, i.e. for $T$ approaching $T_c$ from below.
Like for the GBM choice (\ref{FQs}), 
such a situation can exist for   $ \sigma(T_c) = 0$ and $\frac{3}{2} < \tau \le  2 $. 
Studying the higher $T$-derivatives  of $s_H(T)$ at $T_c$, one can show 
that  for  $ \sigma(T) \equiv  0$  and  for $(n+1)/n \le \tau < n/(n-1)$ ($n=3,4,5,...$) there is a $n^{th}$ order phase  transition
\begin{eqnarray}\label{nth}
& s_H(T_c) = s_Q(T_c),~
s_H^{\prime}(T_c) = s_Q^{\prime}(T_c),~...~ 
 s_H^{(n-1)}(T_c)=  s_Q^{(n-1)}(T_c),~
s_H^{(n)}(T_c) \ne s_Q^{(n)}(T_c),
\end{eqnarray}
 with $ s_H^{(n)}(T_c)=\infty$
for $(n+1)/n < \tau < n/(n-1)$
and  with a finite value of $s_H^{(n)}(T_c)$ for $\tau =(n+1)/n$.\\

\noindent
({\bf II}) The second possibility, $\sigma(T) \equiv 0$, described in the preceding paragraph,  does not give anything new 
compared to the GBM \cite{Goren:81}. 
If the  PT exists, then
the graphical picture of  singularities is basically similar to Fig. 1. The only difference
is that, depending on the PT order,   the derivatives of  $F(s,T) $ function with respect to $s$  should  diverge at $s = s_Q(T_c)$.\\

\noindent
({\bf III})  A principally new possibility exists for $T > T_{cep}$, where $\sigma(T) < 0$.
In this case there exists a  cross-over, if for  $T \le T_{cep}$ the  rightmost 
singularity is $s_H(T)$, which corresponds to the leftmost curve in the left panel of  Fig. 1. 
Under the latter, its existence can be shown as follows. 
Let us solve the equation for singularities (\ref{s*vdw}) graphically (see the right panel of Fig. 1). 
For  $\sigma(T) < 0$ the function $F_Q(s,T)$ diverges at $s = s_Q(T)$. 
On the other hand, the partial derivatives $\frac{\partial F_H(s,T)}{\partial s} < 0$ and 
$\frac{\partial F_Q(s,T)}{\partial s} < 0$ are always negative. Therefore, the  function 
$F(s,T) \equiv  F_H(s,T) +  F_Q(s,T)$ is a monotonically decreasing function of 
$s$, which vanishes at $s \rightarrow \infty$. Since the left hand side of Eq.  (\ref{s*vdw})  is a  monotonically increasing function of $s$, then there can exist a single intersection 
$s^*$ of $s$ and $F(s,T)$ functions. Moreover, for finite  $s_Q(T)$ values this 
intersection can occur  on 
the right hand side of the point $s = s_Q(T)$, i.e.  $s^* > s_Q(T)$ (see the right panel of  Fig. 1). 
Thus, in this case the essential singularity $s = s_Q(T)$ can become the rightmost one
for infinite temperature only.  In other words, the pressure of the pure QGP can be reached 
at infinite $T$, whereas for finite $T$ the hadronic mass spectrum gives a non-zero  contribution into all thermodynamic functions. 
Note that such a behavior is typical for the lattice QCD data at zero baryonic chemical 
potential \cite{Karsch:03}.

In terms of the present model  it is clear that  a cross-over  existence means
a  fast transition of energy or entropy density in a narrow $T$ region  from a dominance 
of  the discrete mass-volume spectrum of light hadrons  to
a dominance of  the  continuous spectrum of heavy QGP bags.  This is exactly the case for  $\sigma(T) < 0$ because  in the right  vicinity of the point $s = s_Q(T)$ the function $F(s,T)$ decreases very  fast and then it gradually decreases as function of $s$-variable. Since,  $F_Q(s,T) $  changes  fast from $F(s,T) \sim F_Q (s,T) \sim s_Q(T)$ to 
$F(s,T) \sim F_H (s,T) \sim s_H(T)$,  their $s$-derivatives should change fast as well. Now, recalling that the change from  $F(s,T) \sim F_Q (s,T)$ behavior to $F(s,T) \sim F_H (s,T)$ in $s$-variable 
corresponds to the cooling of the system (see the right panel of Fig. 1), I conclude that
that there exists a narrow region of temperatures, where the $T$ derivative of    system pressure, i.e. the entropy density, drops down from $\frac{\partial p}{\partial T}  \sim  s_Q(T) + T \frac{d s_Q(T)}{d T} $ to  
$\frac{\partial p}{\partial T}  \sim  s_H(T) + T \frac{d s_H(T)}{d T} $
very fast  compared  to other regions of $T$, if system cools.
If, however, in the vicinity of $T= T_{cep} -0$ the rightmost singularity is $s_Q(T)$, 
then for $T > T_{cep}$  the situation is different  and the cross-over does not  exist. A detailed analysis of
this situation is given in Sect. 4.

Note also that all these nice properties would vanish, if  the reduced surface tension coefficient is  zero or positive above $T_{cep}$. This is  one of the crucial points of the present model which puts forward certain doubts about the vanishing of 
the reduced  surface tension coefficient  in the FDM  and SMM. These doubts are also supported by the first principle results obtained by the Hills and Dales Model  \cite{Bugaev:04b,Bugaev:05a}, because the surface entropy simply  counts the degeneracy of a cluster of a fixed volume and  it  does not physically affect  the surface energy of this cluster.

\section{Generalization to Non-Zero Baryonic Densities}

The possibilities  ({\bf I})-({\bf III}) discussed in the preceding section 
remain unchanged for non-zero baryonic numbers. The latter should be 
included into consideration  to make our model more realistic. To keep 
the presentation simple, I do not consider   strangeness. 
The inclusion of the baryonic charge of the quark-gluon bags 
does not change the two types of singularities of the isobaric partition (\ref{Zs})
and the corresponding equation for them (\ref{s*vdw}), but it 
leads to 
the following modifications of the $F_H$ and $F_Q$ functions:
\begin{eqnarray}\label{FHTmu}
F_H&(s,T,\mu_B)= \sum_{j=1}^n g_j     e^{\frac{b_j \mu_B}{T} -v_js} \phi(T,m_j)\,,
\\
F_Q &(s,T,\mu_B) = {\textstyle u(T, {\mu_B})}
 \int\limits_{V_0}^{\infty}dv~ \frac{ \exp\left[\left(s_Q(T,\mu_B)-s\right)v - \sigma(T) 
v^{\kappa}\right] }{v^{\tau}}\,. 
\label{FQTmu}   
\end{eqnarray}
Here the baryonic chemical potential is denoted as $\mu_B$,  the baryonic charge of 
the $j$-th hadron in the discrete part of the spectrum is $b_j$. The continuous part 
of the spectrum, $F_Q$ can be obtained from  some spectrum $\rho(m,v, b)$
in the spirit of Ref. \cite{Goren:82}, but this is not the aim of the present work. 

The QGP pressure $p_Q = T s_Q(T,\mu_B)$ can be also chosen in several ways. 
Here I  use the bag model pressure 
\begin{eqnarray}\label{sQB}
&\hspace*{-0.2cm}p_Q = \frac{\pi^2}{90}T^4 \left[
 \frac{95}{2} +
\frac{10}{\pi^2} \left(\frac{\mu_B}{T}\right)^2 + \frac{5}{9\pi^4}
\left(\frac{\mu_B}{T}\right)^4 \right]
- B \,, 
\end{eqnarray}
but the more complicated model pressures, even with the
PT of other kind like the transition between the color superconducting QGP 
and the usual QGP, can be, in principle,  used. 

It can be shown \cite{Bugaev:07}  that the sufficient conditions for a PT  existence are  
\begin{eqnarray}\label{SufCondI}
&\hspace*{-0.25cm}
{\textstyle F((s_Q(T,\mu_B\hspace*{-0.05cm}=\hspace*{-0.05cm}0)\hspace*{-0.05cm}+\hspace*{-0.05cm}0),  T,\mu_B=0)  >  s_Q(T,\mu_B=0), } \\
&\hspace*{-0.25cm}
F ((s_Q(T,\mu_B )\hspace*{-0.05cm}+\hspace*{-0.05cm}0),  T,\mu_B)  < s_Q(T,\mu_B)\,,  \forall \mu_B > \mu_A.
\label{SufCondII}
\end{eqnarray}
The  condition (\ref{SufCondI})  provides that the simple pole singularity 
$s^* = s_H(T,\mu_B=0)$ is the rightmost 
one at vanishing $\mu_B=0$ and given $T$, whereas  the condition (\ref{SufCondII}) 
ensures that $s^* = s_Q(T,\mu_B)$ is the rightmost singularity of the isobaric partition for 
all values of the baryonic chemical potential above some positive  constant $\mu_A$. 
This can be seen in Fig. 1 for  $\mu_B$ being a variable.  
Since  $F (s, T,\mu_B)$, where it exists,  is a continuous function of its  parameters,
one concludes that, if the conditions (\ref{SufCondI}) and (\ref{SufCondII}), are fulfilled,
then at some chemical potential $\mu_B^c (T)$ the both singularities should  
be equal. Thus, one arrives at the Gibbs criterion (\ref{PTI}), but for two variables
\begin{eqnarray}\label{PTII}
&  s_H (T, \mu_B^c(T))  =  s_Q (T, \mu_B^c(T)) \,.
 \end{eqnarray}

\noindent 
It is easy to see that the  inequalities  (\ref{SufCondI}) and (\ref{SufCondII}) are  the  sufficient conditions  of a PT existence
for  more complicated functional dependencies of $F_H (s,  T,\mu_B)$ and 
$F_Q (s,  T,\mu_B)$ than the ones used here.

For the  choice (\ref{FHTmu}), (\ref{FQTmu}) and (\ref{sQB})  of 
$F_H (s,  T,\mu_B)$ and $F_Q (s,  T,\mu_B)$ functions the  PT exists at 
$T < T_{cep}$,  because the sufficient conditions (\ref{SufCondI}) and (\ref{SufCondII}) 
can be  easily fulfilled   by a proper choice of the bag constant $B$ and the function 
${\textstyle u(T, \mu_B)} > 0$ for the interval $T \le T_{up}$ with the constant $T_{up} > T_{cep}$. 
Clearly, this is the  1$^{st}$ order PT, since the surface
tension is finite and it provides the convergence of the  integrals (\ref{G1}) and  (\ref{KQ})
in the expression (\ref{sHprime1}), where  the  usual $T$-derivatives should be now 
understood as the partial ones for $\mu_B = const$.  

%
%
\begin{figure}[ht]
\centerline{\hspace*{-0.0cm}\epsfig{figure=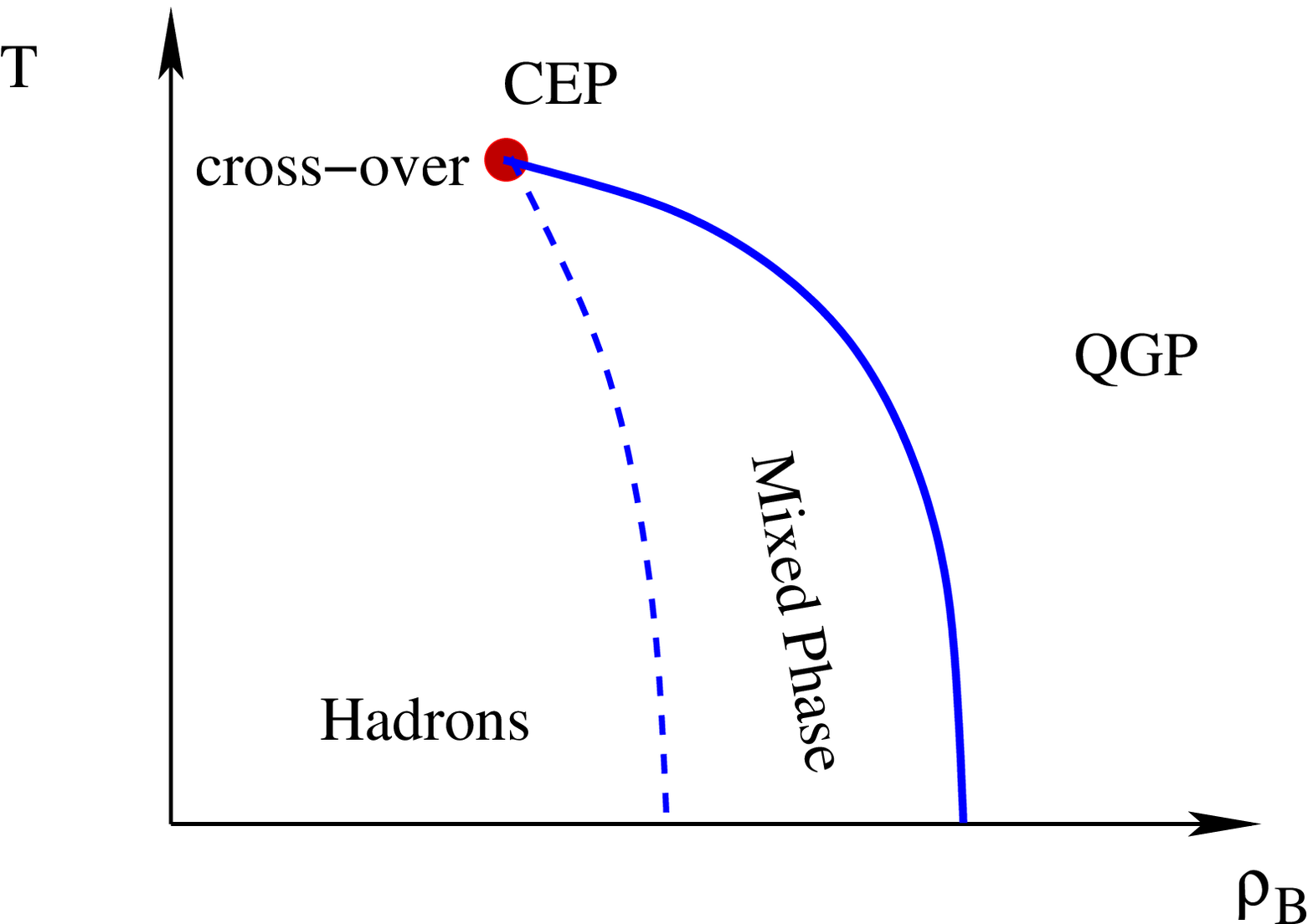,height=5.0cm,width=6.30cm} 
\hspace*{0.4cm}  
\epsfig{figure=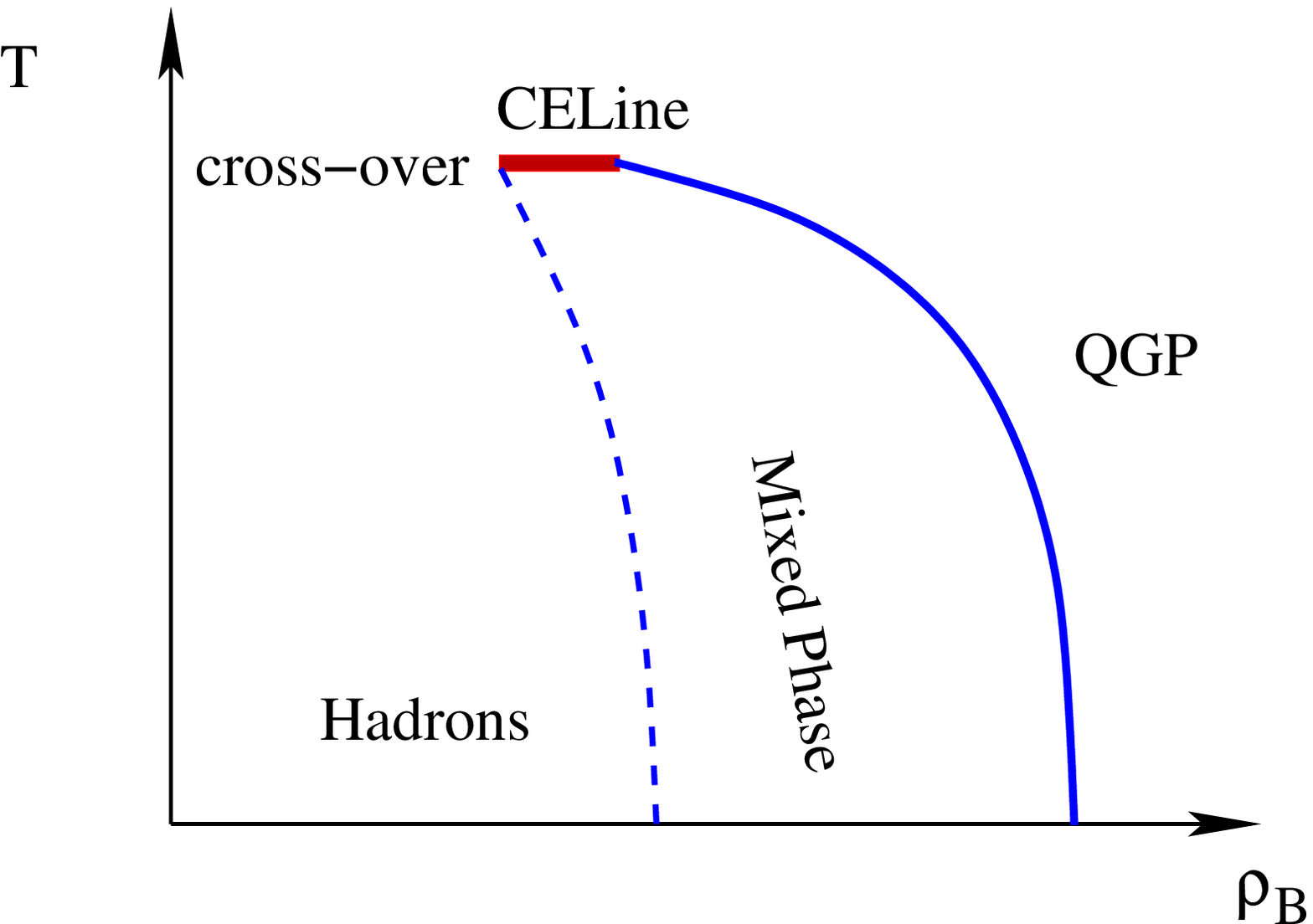,height=5.0cm,width=6.30cm}
}

\vspace*{-0.5cm}
\caption{
\label{fig3}
{\bf Left panel.}
A schematic picture of the deconfinement phase transition diagram 
in the plane of baryonic density $\rho_B$  and  $T$ for 
the 2$^{nd}$ order PT at the critical endpoint (CEP), i.e. for $\frac{3}{2} < \tau \le 2$. 
For the 3$^{rd}$ (or higher) order PT  the  boundary of the mixed 
and hadronic phases (dashed curve) should have the same slope as 
the boundary of the mixed phase and QGP (solid curve) at the CEP. 
\newline
{\bf Right panel.} Same as in the left panel, but for $ \tau >  2$.
The critical endpoint in the $\mu_B-T$ plane generates 
the critical end line (CELine) in the $\rho_B-T$ plane shown by the thick horizontal line. 
This occurs because of the discontinuity of the partial derivatives of $s_H$ and $s_Q$ 
functions with respect to $\mu_B$ and $T$. 
}
\end{figure}

\vspace*{0.7cm}

Assuming that the conditions (\ref{SufCondI}) and (\ref{SufCondII}) 
 are fulfilled by the correct choice of the  model parameters $B$ and  
 ${\textstyle u(T, \mu_B)} > 0$,   one  can see now that at $T = T_{cep}$ there exists 
a PT as well, but its order is defined by the value of $\tau$. As was discussed in the 
preceding section for  $\frac{3}{2} < \tau \le 2$ there  exists the 2$^{nd}$ order PT. 
For  $1 < \tau \le \frac{3}{2}$  there  exist  the  PT of higher order, defined 
by the conditions formulated in Eq.  (\ref{nth}). 
This is a new possibility, which, to my  best knowledge,  does not contradict  to any 
general physical principle (see the left panel in Fig. 2). 

The case $ \tau > 2$ can be ruled out because  there must  exist
the first order PT for  $T \ge T_{cep}$, whereas for  $T < T_{cep}$ there exists 
the cross-over. Thus,  the critical endpoint in $T-\mu_B$ plane  will correspond to 
the critical  interval  in the   temperature-baryonic density plane.  
Since such a structure 
of the  phase diagram in the variables temperature-density  has, to my knowledge,
never  been observed, I conclude that  the case  $ \tau > 2$ is unrealistic (see the right panel in Fig. 2). 
Note that a similar phase diagram exists in the FDM with the only  difference  that 
the boundary of the mixed and liquid phases (the latter in the QGBST model corresponds 
to QGP) is moved to infinite particle density.

\section{Surface Tension Induced Phase Transition}

Using our results for the  case ({\bf III}) of the preceding section, we conclude that 
above $T_{cep}$ there is a cross-over, i.e. the QGP and hadrons coexist together 
up to the infinite values of $T$ and/or $\mu_B$. Now, however, it is necessary to  answer
the question: How can the two different sets of singularities  that exist on two sides of 
the line $T =  T_{cep}$ provide the continuity of the solution of Eq. (\ref{s*vdw})?

It is easy to  answer  this question for $\mu_B < \mu_B^c(T_{cep})$  because 
in this case all partial $T$ derivatives of  $s_H(T, \mu_B) $, which is the rightmost singularity, 
exist and are finite at any point of the line $T =  T_{cep}$.  This can be seen from the
fact that  for the considered  region of parameters  $s_H(T, \mu_B) $ is the rightmost singularity and, consequently, $s_H(T, \mu_B) > s_Q (T, \mu_B)$. The latter inequality 
provides the existence and finiteness of the volume integral in $F_Q(s,T,\mu_B)$.
In combination with  the power $T$ dependence of  the reduced surface tension 
coefficient $\sigma(T)$
the same inequality provides   the existence and finiteness of 
all its partial  $T$ derivatives of $F_Q(s,T,\mu_B)$ regardless  to the sign of 
$\sigma(T)$.  Thus, using the Taylor expansion in powers of $(T - T_{cep})$ 
at any point of the interval
$T =  T_{cep}$ and $\mu_B < \mu_B^c(T_{cep})$, one can calculate 
$s_H(T , \mu_B)$ for the values of  $T > T_{cep}$ which are  inside the convergency  radius of  the Taylor expansion.  

The other situation is for $\mu_B \ge \mu_B^c(T_{cep})$ and $T > T_{cep}$,
namely
in this case above 
the deconfinement PT there must exist a weaker PT 
induced by the disappearance of the reduced  surface tension coefficient. 
To demonstrate this we have solve Eq. (\ref{s*vdw}) in the limit, when $T$ approaches 
the curve $T= T_{cep}$ from above, i.e. for  $T \rightarrow  T_{cep}+0$, and study the 
behavior of $T$ derivatives of the  solution of Eq. (\ref{s*vdw}) $s^*$ for fixed values of 
$\mu_B$.
For this purpose  we have to evaluate the  integrals ${\cal K}_\tau (\Delta,\gamma^2)$
introduced in Eq. (\ref{KQ}). 
Here  the notations 
$\Delta \equiv s^* - s_Q(T, \mu_B)$ and $\gamma^2 \equiv - \sigma (T) > 0$ are  introduced for convenience. 

To avoid the unpleasant behavior for $\tau \le 2$ it is convenient to transform  (\ref{KQ}) 
further on by integrating by parts: 
\begin{eqnarray}\label{KQ2}
{\cal K}_\tau (\Delta,\gamma^2) \, \equiv & ~ g_\tau(V_0)  - \frac{\Delta}{(\tau-1)} {\cal K}_{\tau -1} (\Delta,\gamma^2) +
\frac{\kappa \, \gamma^2}{(\tau-1)} {\cal K}_{\tau-\kappa} (\Delta,\gamma^2) \,,
\end{eqnarray}
where the regular function $g_\tau(V_0) $ is defined as
\begin{eqnarray}\label{gtau}
&  g_\tau(V_0) \equiv  \frac{1}{(\tau-1)\, V_0^{\tau-1}} \exp\left[ -\Delta V_0 + 
\gamma^2 V_0^{\kappa}\right] \,. 
 \end{eqnarray}
For $\tau - a > 1$ one can change the variable of integration 
$v \rightarrow z / \Delta$ and  rewrite $ {\cal K}_{\tau-a} (\Delta,\gamma^2)$ as 
\begin{eqnarray}
\hspace*{-0.25cm}
 {\cal K}_{\tau- a} (\Delta,\gamma^2)  & = & \Delta^{\tau- a-1} \hspace*{-0.15cm} 
 \int\limits_{V_0 \Delta}^\infty  \hspace*{-0.15cm}
dz ~\frac{\exp\left[- z  + \frac{\gamma^2}{\Delta^\kappa} z^{\kappa} 
\right] }{z^{\tau- a}}   \equiv 
%
\Delta^{\tau-a-1} \, {\cal K}_{\tau-a} \left(1, \gamma^2\Delta^{-\kappa} \right) \,.
\label{KQ3}
 \end{eqnarray}
This result shows that  in the limit $\gamma \rightarrow 0$, when the rightmost 
singularity must  approach $s_Q(T,\mu_B)$ from above, i.e. $\Delta \rightarrow 0^+$, the function (\ref{KQ3}) behaves as  $ {\cal K}_{\tau- a} (\Delta,\gamma^2) \sim \Delta^{\tau-a-1} + O(\Delta^{\tau-a})$. This is so because for $\gamma \rightarrow 0$ 
the ratio $\gamma^2\Delta^{-\kappa}$ cannot go to infinity, otherwise  the function 
${\cal K}_{\tau-1} \left(1, \gamma^2\Delta^{-\kappa} \right) $,  which enters into the right hand side of  (\ref{KQ2}),   would diverge exponentially 
and this   makes impossible an existence of the solution of Eq. (\ref{s*vdw}) for 
$T = T_{cep}$. The analysis shows that for $\gamma \rightarrow 0$  there exist  two possibilities: either  $\nu \equiv \gamma^2\Delta^{-\kappa} \rightarrow Const$ or 
$\nu \equiv \gamma^2\Delta^{-\kappa} \rightarrow 0$.
The most straightforward  way to analyze these possibilities for  $\gamma \rightarrow 0$ is to assume the following behavior 
\begin{eqnarray} \label{Dasgamma}
&  \Delta  =  A\, \gamma^\alpha +  O(\gamma^{\alpha+1}) \,,~~ \Rightarrow~~
\frac{ \partial \Delta}{\partial T}   =   \frac{ \partial \gamma}{\partial T} 
 \left[A\,\alpha\,  \gamma^{\alpha-1} +  O(\gamma^{\alpha})\right] 
 \sim \frac{(2\,k +1) A \,\alpha\,  \gamma^{\alpha}}{2\, (T - T_{cep}) },
%
\end{eqnarray}
and find out the $\alpha$ value by equating the $T$ derivative of $\Delta$ with  the $T $ derivative (\ref{sHprime1}).

The analysis shows \cite{Bugaev:07}  that for  $\Delta^{2-\tau} \le  \gamma \gamma^\prime  \Delta^{1-\kappa}$  one finds
\begin{eqnarray}\label{alpha1}
&  
%
\gamma^{\alpha -2} \sim \Delta^{1-\kappa} \Rightarrow ~ \alpha \kappa = 2 ~~{\rm for}~~ \tau \le 1 + \frac{\kappa}{2 k + 1} \,. 
 \end{eqnarray}

Similarly, for  $\Delta^{2-\tau} \ge  \gamma \gamma^\prime  \Delta^{1-\kappa}$ one 
obtains $\gamma^{\alpha -1} \gamma^\prime \sim \Delta^{2 -\tau}$ and, consequently,
\begin{eqnarray}\label{alpha2}
& 
\alpha  = \frac{2}{(\tau-1)(2k+1)} ~~{\rm for}~~ \tau \ge 1 + \frac{\kappa}{2 k + 1} \,.
 \end{eqnarray}
%

%
%
\begin{figure}[ht]
\centerline{\epsfig{figure=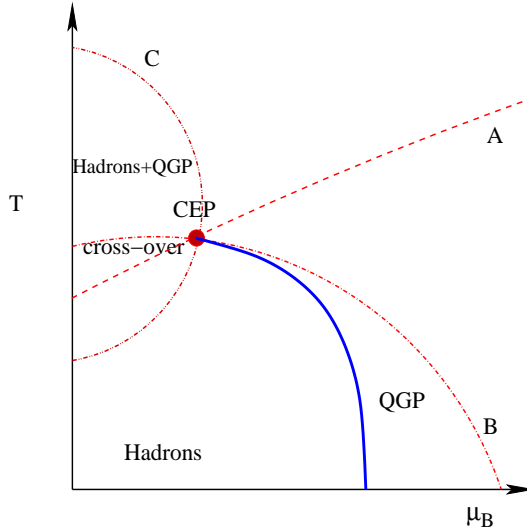,height=7.cm,width=7.0cm}}
\vspace*{-0.5cm}
\caption{
A schematic picture of the deconfinement phase transition diagram (full curve)
in the plane of  baryonic chemical potential $\mu_B$  and  $T$ for 
the 2$^{nd}$ order PT at the tricritical endpoint (CEP).
The model predicts an existence of the  surface induced PT of the 2$^{nd}$ or
higher order (depending on the model parameters). This PT starts at the CEP and goes 
to higher values of $T$ and/or $\mu_B$. Here it  is shown by the dashed curve CEP-A, if the phase diagram is endless, or
by the dashed-dot curve CEP-B, if the phase diagram ends at $T= 0$, or by 
the dashed-double-dot curve CEP-C,  if the phase diagram ends at $\mu_B= 0$.
Below (above) each of A or B curves the reduced surface tension coefficient 
is positive (negative).  For  the curve C the surface tension coefficient is positive 
outside of it. 
}
\end{figure}

\vspace*{0.7cm}

Summarizing these results for $\gamma \rightarrow 0$,
one can write the expression for the second derivative of $\Delta$  as \cite{Bugaev:07}: 
\vspace*{-0.0cm}
\begin{eqnarray}\label{D2sdT2tot}
\frac{ \partial^2 \Delta}{\partial T^2}   \hspace*{-0.0cm}  \sim 
\left\{
\begin{tabular}{ll}
\vspace{0.1cm}  ${\textstyle \left[ \frac{T-T_{cep}}{T_{cep}} \right]}^{\frac{2k+1}{\kappa}-2} $\,, &  \hspace*{-0.2cm} $ \tau \le 1 + \frac{\kappa}{2 k + 1}$\,, \\
& \\
${\textstyle \left[ \frac{T-T_{cep}}{T_{cep}} \right]}^{\frac{3-2\tau}{\tau-1} } $\,, &  \hspace*{-0.2cm} $\tau \ge 1 + \frac{\kappa}{2 k + 1} $\,.
\end{tabular}
\right.  \hspace*{-0.3cm}
\end{eqnarray}
The last result shows us that, depending on $\kappa$ and $k$ values, 
the second derivatives of $s*$ and $s_Q(T,\mu_B)$ can differ from each other  for 
$ \frac{3}{2} < \tau < 2$ or can be equal for $ 1 < \tau \le \frac{3}{2}$.
In other words, it is found  that at  the line $T = T_{cep}$ there exists  the 2$^{nd}$ order PT for $ \frac{3}{2} < \tau < 2$ and the higher order PT   for $ 1 < \tau \le  \frac{3}{2}$,
which separates the pure QGP phase from the region  of  a cross-over, i.e.  the mixed states of hadronic and QGP bags. Since it exists at the line of a zero surface tension, 
this PT  will be called the {\it surface induced PT.} 
For instance,  from (\ref{D2sdT2tot}) it follows that for $k = 0$ and  $\kappa > \frac{1}{2}$ there is  the 2$^{nd}$ order PT, whereas  for   $k = 0$ and  $\kappa = \frac{1}{2}$ or for $k > 0$ and  
$\kappa < 1 $ there is  the 3$^{d}$ order PT, and so on.

\vspace*{-0.0cm}

Since the analysis performed in  the present section did not include any $\mu_B$ derivatives 
of $\Delta$, it remains valid for   the $\mu_B$ dependence of the 
reduced surface tension coefficient, i.e. for $T_{cep} (\mu_B)$.
Only it is  necessary to make a few comments on a possible location of  
the {\it surface tension null line}  $T_{cep} (\mu_B)$.  
In principle, such a null line can be located anywhere, if its location does not contradict 
to the sufficient conditions (\ref{SufCondI}) and (\ref{SufCondII}) of   the  1$^{st}$ deconfinement PT existence.  Thus, the surface tension null line must cross the 
deconfinement line in the $\mu_B-T$ plane at a single point which is  the tricritical endpoint
$(\mu_B^{cep}; T_{cep} (\mu_B^{cep})) $, whereas for  $\mu_B > \mu_B^{cep} $ the 
null line should have higher temperature for the same $\mu_B$   than the deconfinement one, i.e. 
$T_{cep} (\mu_B) > T_c  (\mu_B) $ (see Fig. 3).  Clearly, there exist  two distinct cases
for the surface tension null line: either it is endless, or it ends at 
zero temperature or at other singularity, like  the Color-Flavor-Locked  phase. 
From the present lattice QCD data \cite{Karsch:03} it follows that the case C in Fig. 3 is the least possible.

To understand the meaning of the  surface induced PT it is instructive to
 quantify the difference between phases by looking into the mean size of the bag:
\begin{eqnarray}\label{BagSize}
& 
\langle v \rangle \equiv - \frac{\partial \ln F(s, T, \mu_B)}{\partial~ ~s} \biggl|_{s = s^*-0} \,.
 \end{eqnarray}
As was shown 
in hadronic phase phase $\Delta > 0$ and, hence, it consists of the bags of finite mean volumes, whereas, 
by construction, the QGP phase is a single infinite bag. For the cross-over states 
$\Delta > 0$ and, therefore, they are the bags of finite mean volumes, which 
gradually increase, if  the rightmost singularity approaches $s_Q(T,\mu_B)$, i.e.
at very large values $T$ and/or $\mu_B$. Such a classification is useful 
to distinguish  QCD phases of present model: it shows that hadronic and cross-over 
states are separated from the QGP phase by the 1$^{st}$ order deconfinement PT and 
by the 2$^{nd}$ or higher order PT, respectively.

\section{Concluding Remarks}

Here I presented an analytically solvable  statistical model  which simultaneously  describes 
the 1$^{st}$ and 2$^{nd}$ order PTs with a cross-over. The approach is general and can be used 
for  more complicated parameterizations of  the hadronic mass-volume spectrum, if in the vicinity of
the deconfinement PT region the discrete and continuous parts  of this spectrum  can be expressed in the form of Eqs. (\ref{FHTmu}) and  (\ref{FQTmu}), respectively. Also the actual parameterization of 
the QGP pressure $p = T s_Q(T,\mu_B)$ was not used so far, which means that our result can be extended to more complicated functions, that can contain other phase transformations (chiral PT,
or the PT to color superconducting phase)   provided that the sufficient  conditions (\ref{SufCondI}) and (\ref{SufCondII})  for the deconfinement PT existence  are satisfied. 

In this model the desired properties of the deconfinement phase diagram are achieved by accounting for the temperature dependent surface tension of the quark-gluon bags. 
As was shown, it is crucial for the cross-over existence that at  $T= T_{cep}$ 
the reduced surface tension 
coefficient vanishes and remains negative for temperatures above  $T_{cep}$.
Then the deconfinement $\mu_B-T$  phase diagram has the 1$^{st}$ PT  at  
$\mu_B > \mu^c_B( T_{cep})$ for 
$ \frac{3}{2} < \tau <  2$ , which degenerates into the 2$^{nd}$ order PT 
(or higher order PT for  $ \frac{3}{2} \ge \tau >1 $)  at 
$\mu_B = \mu^c_B( T_{cep})$, and a cross-over for $0 \le \mu_B < \mu^c_B( T_{cep})$.
These two ingredients  drastically change the critical properties of the GBM 
\cite{Goren:81} and resolve the long standing problem of  a unified description of  the 
1$^{st}$ and 2$^{nd}$ order PTs and  a cross-over,
which, despite all claims,  was not resolved in Ref. \cite{Goren:05}. 
In addition, it was found that at the null line of the surface tension there must exist 
the surface induced  PT of the 2$^{nd}$ or higher order, which separates 
the pure QGP from the mixed states of hadrons and QGP bags, that coexist above
the cross-over region (see Fig. 3).  Thus, the QGBST model predicts that the QCD critical endpoint 
is the tricritical endpoint. It would be interesting  to verify this prediction
with the help of the lattice QCD analysis. 
For this one will need to study the behavior of the bulk and surface contributions to 
the free energy of the  QGP bags  and/or  the string connecting 
the static quark-antiquark pair.

However,  the QGP bags created in the experiments  have finite mass, volume, life-time  and, hence,  the strong discontinuities  which are typical for the 1$^{st}$ order  PT should be smeared out 
which would  make them hardly distinguishable from the cross-over. Thus, to seriously discuss 
the signals of the 1$^{st}$  order deconfinement PT and/or  the tricritical endpoint,  one needs to 
solve the  finite volume version of the QGBST model like  it was done for the SMM \cite{Bugaev:04a} and the GBM \cite{Bugaev:05c,Bugaev:07b}.  This, however, is not sufficient because, in order to make any reliable prediction for experiments, the finite volume equation of state must be used  in   hydrodynamic equations which, unfortunately, are not suited for such a  purpose.  Thus, we are facing a necessity to return to the foundations of heavy ion phenomenology and to modify them according to the requirements of the experiments. 

In addition, to apply the QGBST model to the experiments it is nesseary to make it more realistic:
it seems that for the mixture of hadrons and QGP bags above  the cross-over  line
it is necessary to include the relativistic treatment of hard core repulsion 
\cite{Rvdw:1,Rvdw:2} for lightest hardons and to include into statistical description 
the medium dependent width of  hadronic resonances and QGP bags,
which, as argued  in Ref.  \cite{Blaschke:03},  may completely  change our  
understanding  of the cross-over mechanism.

\bigskip

{\bf Acknowledgments.} The author thanks  V. K. Petrov for  valuable comments.  The research made in this work 
was supported  by the Program ``Fundamental Properties of Physical Systems 
under Extreme Conditions''  of the Bureau of the Section of Physics and Astronomy  of
the National Academy of Science of Ukraine.


%

%
%
\end{document}